\RequirePackage{fix-cm} 
\documentclass[a4paper, twoside, reqno, 12pt, dvips]{amsart}

\usepackage{fixltx2e}     

\usepackage{amssymb}
\usepackage{amsmath}
\usepackage{mathrsfs, dsfont}

\usepackage{a4wide}

\usepackage{indentfirst}
\usepackage{graphicx}
\usepackage{psfrag}

\usepackage[usenames,dvipsnames]{pstricks}
\usepackage{epsfig}
\usepackage{pst-grad} 
\usepackage{pst-plot} 
\usepackage{infix-RPN}

\usepackage{subfigure}

\usepackage{acronym}
\usepackage{longtable}

\usepackage[english]{babel}
\usepackage[latin1]{inputenc}

\usepackage{color}
\definecolor{oneblue}{rgb}{0,0.0,0.75}
\usepackage[colorlinks,
            urlcolor=oneblue,
            linkcolor=oneblue,
            citecolor=oneblue,
            bookmarksopen=false,
            pagebackref]{hyperref}

\numberwithin{equation}{section}

\newcommand{\R}{\mathbb{R}}
\newcommand{\E}{\mathbb{E}}
\newcommand{\I}{\mathcal{I}}
\newcommand{\C}{\mathcal{C}}
\newcommand{\T}{\mathcal{T}}
\newcommand{\N}{\mathcal{N}}
\renewcommand{\S}{\mathcal{S}}
\renewcommand{\H}{\mathcal{H}}
\newcommand{\xxi}{\boldsymbol{\xi}}
\newcommand{\ttheta}{\boldsymbol{\theta}}

\begin{document}

\title{Long wave runup on random beaches}

\author[D. Dutykh]{Denys Dutykh$^*$}
\address{LAMA, UMR 5127 CNRS, Universit\'e de Savoie, Campus Scientifique, 73376 Le Bourget-du-Lac Cedex, France}
\email{Denys.Dutykh@univ-savoie.fr}
\urladdr{http://www.lama.univ-savoie.fr/~dutykh/}
\thanks{$^*$ Corresponding author}

\author[C. Labart]{C\'eline Labart}
\address{LAMA, UMR 5127 CNRS, Universit\'e de Savoie, Campus Scientifique, 73376 Le Bourget-du-Lac Cedex, France}
\email{Celine.Labart@univ-savoie.fr}
\urladdr{http://www.lama.univ-savoie.fr/~labart/}

\author[D. Mitsotakis]{Dimitrios Mitsotakis}
\address{IMA, University of Minnesota, 114 Lind Hall, 207 Church Street SE,
Minneapolis MN 55455, USA}
\email{dmitsot@gmail.com}
\urladdr{http://sites.google.com/site/dmitsot/}

\begin{abstract}
The estimation of the maximum wave runup height is a problem of practical importance. Most of the analytical and numerical studies are limited to a constant slope plain shore and to the classical Nonlinear Shallow Water (NSW) equations. However, in nature the shore is characterized by some roughness. In order to take into account the effects of the bottom rugosity various ad-hoc friction terms are usually used. In this paper we study the effect of the roughness of the bottom on the maximum runup height. A stochastic model is proposed to describe the bottom irregularity and its effect is quantified using Monte-Carlo simulations. For the discretization of the NSW equations we employ modern finite volume schemes. Moreover, the results of the random bottom model are compared with the more conventional approaches.
\end{abstract}


\maketitle

\tableofcontents

\section{Introduction}

The estimation of the long wave runup on a sloping beach is a practical problem which attracts nowadays a lot of attention due in part to the intensive human activity in coastal areas. The main demand comes from the coastal and civil engineering but also from coastal communities which are exposed to the tsunami wave hazard \cite{Syno2006}. Consequently, a lot of effort is devoted to the development of fast and accurate estimation methods of the wave runup and horizontal excursion over a sloping beach \cite{Tadepalli1996, Kanoglu2006, Didenkulova2008, Madsen2010}. In general this problem is solved in simplified geometries (e.g. constant slope beach) and in the framework of Linear or Nonlinear Shallow Water (LSW, NSW) equations. However, more general situations may require the application of other models and different numerical techniques (see e.g. \cite{LWL, MarcheBonneton2007, Dutykh2009a, Dutykh2010} and the references therein).

In practice, the available data are always subject to some uncertainties. For example, the bathymetry is known only in a discrete number of scattered points, while in reality the shores are characterized by some rugosity. The missing information can be modeled by the inclusion of random effects. These circumstances have lead several authors to consider water wave propagation in random media \cite{Gurevich1993, Bouard2008, Nachbin2010}. In the present study we model the natural beach roughness by small random perturbations of the smooth average bottom profile. The long wave dynamics are described by the classical NSW equations. We note that the dispersive effects could also be included (see \cite{Dutykh2010}), however they do not modify qualitatively the results that follows bellow. The main effect of the dispersion is a small reduction of the maximum runup height due to the wave energy flux to shorter wavelengths.

\section{Mathematical model and results}

Consider an incompressible perfect fluid layer bounded below by the solid bottom $d(x)$ and above by the free surface $\eta(x,t)$. In the present study we are interested in the long wave regime which is described by the NSW equations:
\begin{eqnarray}
  H_t + (Hu)_x &=& 0, \label{eq:nsw1} \\
  (Hu)_t + \bigl(Hu^2+\textstyle{\frac{g}{2}}H^2\bigr)_x &=& gH d_x - gH\S_f, \label{eq:nsw2}
\end{eqnarray}
where $H(x,t) = d(x) + \eta(x,t)$ is the total water depth and $u(x,t)$ is the depth-averaged fluid velocity. The channel bottom $d(x)$ is assumed to be a sloping beach described by the depth function $d(x) = d_0 - \tan\delta\cdot(x+\ell)$, where $\delta$ is the constant bottom slope and $\ell$ is the half-length of the physical domain. Parameters $d_0$, $\ell$, $\delta$ are chosen so that a dry sloping area is below the still water level (see Table \ref{tab:Params}). The term $\S_f$ is included to model some friction effects and it will be taken zero unless otherwise noted. We consider the Boundary Value Problem (BVP) posed on the one-dimensional interval $\I = [-\ell, \ell]$, where on the right boundary $x = \ell$ we impose the so-called wall boundary condition $u(\ell,t) = 0$ (in our simulations the wave front does not achieve this point), while on the left end $x = -\ell$ we generate an incoming wave of height $\eta(-L,t) = -a_0\sin(\omega_0 t)\H(T_0-t)$, where $\H(t)$ is the Heaviside step-function and $T_0 = 2\pi/\omega_0$ is the wave period. In other words, we generate a shorewared traveling, one-period monochromatic leading depression wave. The values of the various physical and numerical parameters used in this study are given in Table \ref{tab:Params}.

\begin{table}
\caption{Various parameters used in this study. Note that with the present choice of parameters $d_0$ and $g$ we solve the governing equations \eqref{eq:nsw1}, \eqref{eq:nsw2} in the dimensionless form.}
\label{tab:Params}
\begin{tabular}{c|c}
\hline\hline
\textrm{Parameter} & \textrm{Value} \\
\hline
Domain half-length, $L$ & 17.0 \\
Bottom slope, $\tan\delta$ & 0.06 \\
Gravity acceleration, $g$ & 1.0 \\
Water depth at the left end, $d_0$ & 1.0 \\
Incoming wave amplitude, $a_0$ & 0.15 \\
Incoming monochromatic wave frequency, $\omega_0$ & 0.2 \\
Number of control volumes, $N$ & 1000 \\
Number of Monte-Carlo runs, $M$ & 1000 \\
\hline\hline
\end{tabular}
\end{table}

\begin{figure}
  \centering
  \includegraphics[width=0.7\textwidth]{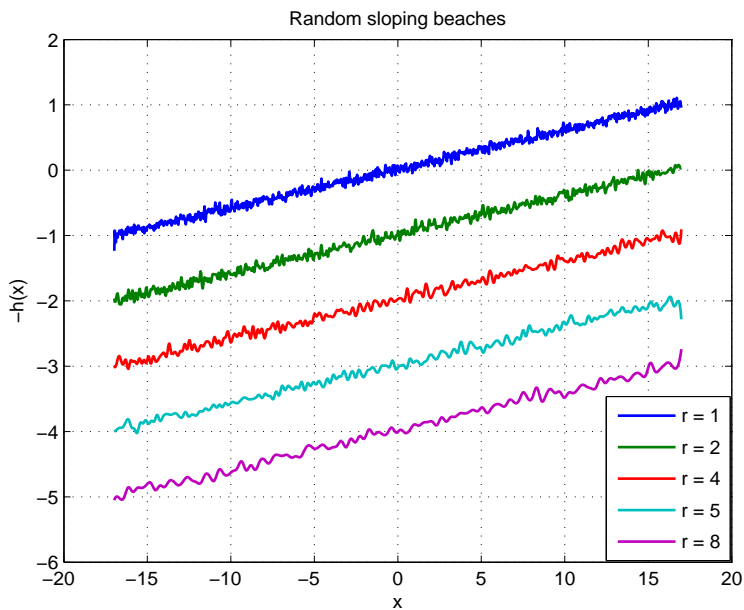}
  \caption{A sample realization of random bottoms for $\sigma = 5\times 10^{-2}$ and various values of the regularity parameter $r = 1$, $2$, $4$, $5$ and $8$ starting correspondingly from the top.}
  \label{fig:randBots}
\end{figure}

\begin{figure}
  \centering
  \includegraphics[width=0.7\textwidth]{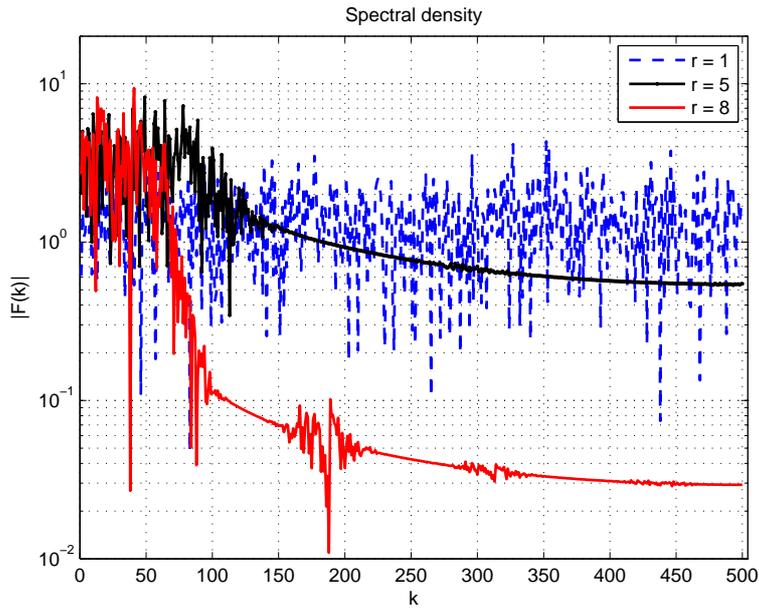}
  \caption{Spectral density of the bottom perturbation for several values of the regularity parameter $r = 1, 2, 8$ (starting from the top).}
  \label{fig:Specdens}
\end{figure}

The interval $\I$ is divided into cells $\C_i = [x_{i-\frac12}, x_{i+\frac12}]$ of length $\Delta x_i = x_{i+\frac12} - x_{i-\frac12}$ and $x_i = \frac12\bigl(x_{i-\frac12} + x_{i+\frac12}\bigr)$ denotes the midpoint of $\C_i$, $i=1,\ldots,N$. Without any loss of generality we assume the partition of cells $\T = \{\C_i\}_{i=1}^{N}$ is uniform. In order to model the bottom roughness we construct a random perturbation in the following way. Let us fix an integer number $r\geq 1$ which will be referred to as the regularity parameter. Then, on each cell $\{\C_{jr}\}_{j=1}^{m} \subset \T$, with $m=\lfloor\frac{N}{r}\rfloor$, we generate a normally distributed pseudorandom variable $\xi_j \sim \N(0,\sigma^2)$, where the parameter $\sigma$ characterizes the perturbation magnitude since $|\xi_j|<1.96\cdot\sigma$ with probability 95\%. Constructed in this way the random vector $\xxi = \{\xi_j\}_{j=1}^{m}$ is interpolated on the whole grid using cubic splines, for example, to obtain a particular realization of micro irregularities. The discrete bathymetry function becomes $d_i = d_0 - \tan\delta\cdot(x_i+\ell) + \xi_i$ on each cell $\C_i$. Several realizations of the random bottom for various values of $r$ are shown on Fig. \ref{fig:randBots}. If $r=1$ we obtain a white noise while increasing this parameter is equivalent to the application of a spectral filtering operation (see Fig. \ref{fig:Specdens}).

The hyperbolic system of NSW equations is discretized using the finite volume method, cf. \cite{Dutykh2010}. Specifically we use the characteristic flux approach, \cite{Ghidaglia2001}, combined with the UNO2 space reconstruction procedure, \cite{HaOs}. The well-balancing of the scheme is achieved by applying the well known hydrostatic reconstruction method \cite{Audusse2005}. The run-up algorithm description can be found in \cite{Dutykh2009a, Dutykh2010}. For the time discretization we use the 3rd order Bogacki-Shampine Runge-Kutta scheme with adaptive time step selection.

\begin{figure}
  \centering
  \includegraphics[width=0.7\textwidth]{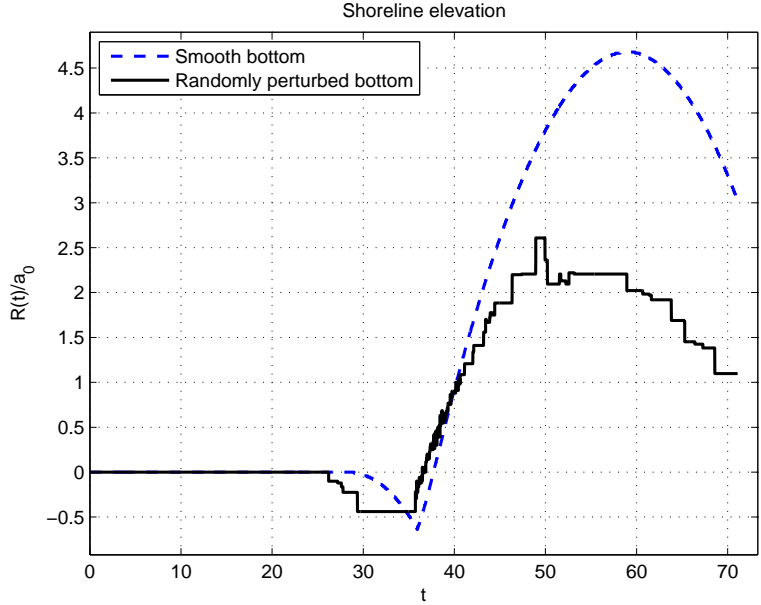}
  \caption{The shoreline motion in the case of a smooth shore (the blue dashed line) and a particular realization of the random bottom with $\sigma = 10^{-2}$, $r=1$ (black solid line).}
  \label{fig:shoreline}
\end{figure}

Once the parameters $\sigma$ and $r$ have been chosen, we can generate a particular realization of the rough sloping beach and solve the BVP to determine the maximum wave runup. The shoreline motion of one particular realization with $\sigma = 10^{-2}$ and $r = 1$ is represented on Fig. \ref{fig:shoreline}. For comparison, the shoreline behavior in the idealized smooth bottom case is also represented on Fig. \ref{fig:shoreline} with the blue dashed line. One can see that the main effect of the bottom rugosity is the reduction of the maximum wave run-up height $R_{\max}$. In this particular simulation the wave run-up has been reduced by a factor of 2 approximately. Sometimes this effect is referred to as the apparent diffusion, cf. \cite{Nachbin2010}. Intuitively we can understand this outcome since a wave dissipates more energy due to the interaction with these micro irregularities.

One of the main questions we address here is to quantify the run-up reduction when the bottom roughness varies. Our approach consists of performing direct numerical simulations of this process over random bottoms instead of adding some ad-hoc terms to model this roughness. We will return to this point below. In probabilistic terms we would like to estimate the expectation $\E(R_{\max})$ over all possible realizations of the random bottom noise.

Since a random bottom perturbation is constructed in the discrete space, the dimension of the random parameters vector $\xxi\in\R^m$ scales with the number of control volumes $N$ in our spatial discretization of the interval $\I$. More precisely $m = \lfloor N/r \rfloor$, where $r\geq 1$ is the noise regularity parameter introduced above. The discrete space in our simulation is of dimension $N$ which is typically of the order $10^3$ (see Table \ref{tab:Params}). This value is imposed by the accuracy requirements of our direct simulations and this rather high dimension is a limiting factor for the choice of the expectation $\E(R_{\max})$ numerical method estimation. Popular nowadays the polynomial chaos expansion method does not apply if the number of random parameters is typically greater than two. The Quasi-Monte-Carlo approach fails for dimensions higher than 200 because of substantial difficulties to generate a low discrepancy sequence of random vectors of such a large dimension. Consequently, we are limited to the standard Monte-Carlo method which is not sensitive to the stochastic problem dimension. However, we can apply a variance reduction method described below.

\begin{figure}
   \centering
   \includegraphics[width=0.75\textwidth]{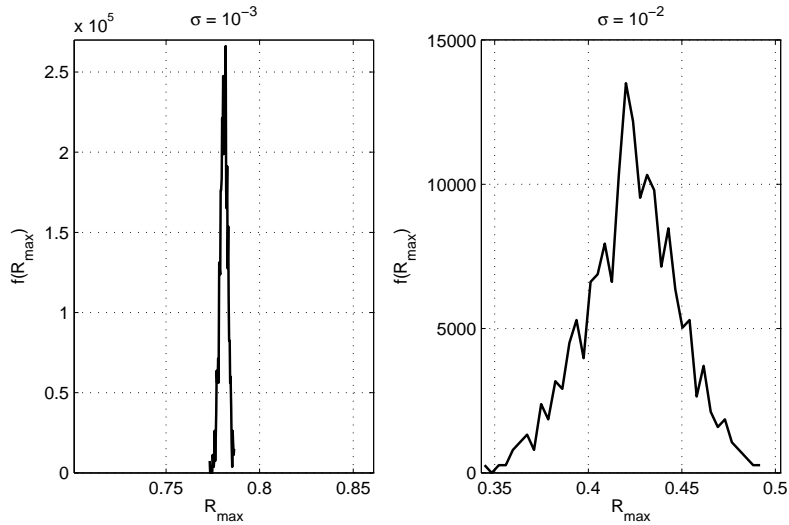}
   \caption{Probability density distribution for $\sigma = 10^{-3}$ (left image, $\E(R_{\max}) = 0.78088$, 5\% and 95\% quantiles are equal to $0.77734$ and $0.78416$ respectively) and $\sigma = 10^{-2}$ (right image, $\E(R_{\max}) = 0.42281$, 5\% and 95\% quantiles are equal to $0.37857$ and $0.46261$ respectively). Note the difference in the vertical scales and in the horizontal extent of two distribution functions.}
   \label{fig:ProbDens}
\end{figure}

In order to estimate $\E(R_{\max})$, we simulate $M$ random bottom realizations, and for each case $j$ we compute numerically the maximum run-up $R^{(j)}_{\max}$. We approximate $\E(R_{\max})$ by the mean $S_M:=\frac{1}{M}\sum_{j=1}^M R^{(j)}_{\max}$. According to the central limit theorem we know that $\E(R_{\max})$ belongs to the interval
$\bigl[S_M-1.96\sqrt{\frac{\sigma^2_M}{M}}, S_M+1.96\sqrt{\frac{\sigma^2_M}{M}}\bigr]$ with a $95\%$ level of confidence, where $\sigma^2_M := \frac{1}{M-1}\sum_{j=1}^M (R^{(j)}_{\max}-S_M)^2$ is an unbiased converging estimator of the variance of $R_{\max}$. To reduce the size of the confidence interval, we can either increase $M$ (which requires more computational time) or try to find a random variable with mean $\E(R_{\max})$ and variance smaller than $\mbox{Var}(R_{\max})$. We opt for
the second possibility -- the so-called variance reduction technique. Since $R_{\max}$
can be seen as a function $R(\xxi)$, where $\xxi$ follows a centered
Gaussian law $\N(\boldsymbol{0}_m,\sigma^2\mathcal{I}_m)$, we can use the adaptive importance sampling technique proposed in \cite{Lapeyre2011}. This method uses the fact that: $\forall \ttheta \in \R^m$, $\E(R(\xxi)) = \E(R(\xxi+\ttheta) e^{-\ttheta \cdot \xxi-\frac{|\ttheta|^2}{2}})$. Then, one can construct an algorithm which finds the parameter vector $\ttheta^{\star}$ minimizing the variance of $H(\ttheta,\xxi) := R(\xxi+\ttheta)e^{-\ttheta\cdot\xxi-\frac{|\ttheta|^2}{2}}$. Then, the average value $\E(R_{\max})$ is approximated by $\overline{S}_M := \frac{1}{M}\sum_{j=1}^M H(\ttheta_{j-1},\xxi_j)$, where $\{\ttheta_j\}_{j=1}^M$ is a sequence converging to $\ttheta^{\star}$. We refer to \cite[Section 2.2]{Lapeyre2011} for theoretical results on the central limit theorem in this adaptive case where the random variables are not independent anymore. This algorithm allows us to reduce the variance by a factor of two approximately. In our computations the confidence interval length has never exceeded 0.5\% of the corresponding maximum run-up value with parameter $M$ specified in Table \ref{tab:Params}. The probability density function of the $R_{\max}$ distribution for $r=1$ and two values of $\sigma$ ($10^{-3}$ and $10^{-2}$) are depicted on Figure \ref{fig:ProbDens}.

\begin{figure}
  \centering
  \includegraphics[width=0.75\textwidth]{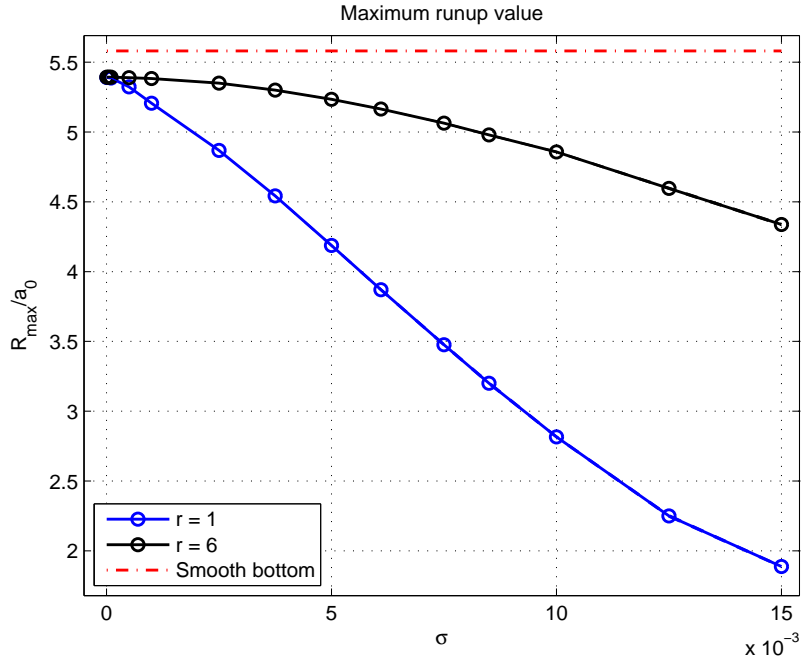}
  \caption{Maximum runup value as a function of the perturbation characteristic magnitude $\sigma$ for the irregular case $r=1$ (blue line) and the regularized noise $r=6$ (black line). For comparison, the red dash-dotted line represents the maximum runup value for the smooth bottom case.}
  \label{fig:Rsigma}
\end{figure}

\begin{figure}
  \centering
  \includegraphics[width=0.75\textwidth]{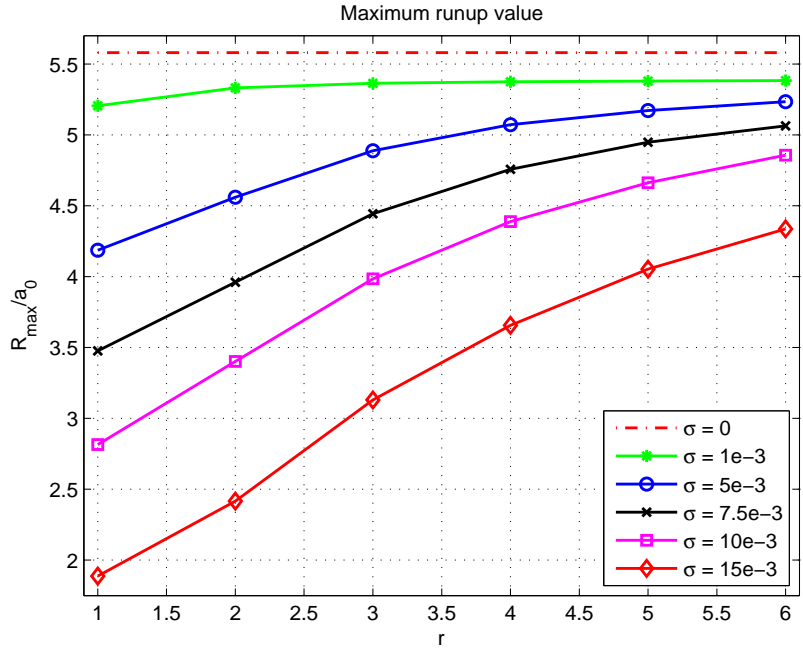}
  \caption{The maximum runup value as a function of the regularity parameter $r$ for several values of the perturbation magnitude $\sigma$ (increasing from the top).}
  \label{fig:Rreg}
\end{figure}

The Monte-Carlo simulation results are presented on Figs. \ref{fig:Rsigma} and \ref{fig:Rreg}. The dependence of the maximum run-up $R_{\max}$ value on the roughness magnitude $\sigma$ for two fixed values of the noise regularity $r = 1$ and $6$ is shown on Fig. \ref{fig:Rsigma}. On the other hand, the dependence of $R_{\max}$ on the regularity parameter $r$ for several fixed values of $\sigma$ is represented on Fig. \ref{fig:Rreg}. We can see that the bottom roughness reduces significantly the wave run-up height while the noise regularization has an antagonistic effect.

\begin{figure}
  \centering
  \includegraphics[width=0.75\textwidth]{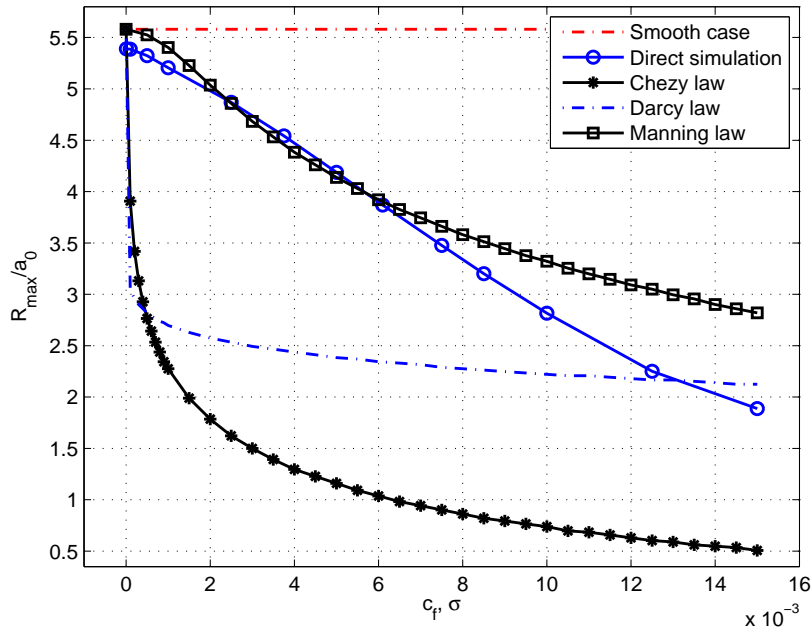}
  \caption{Comparison of the runup reduction effect for various ad-hoc friction terms and the random bottom perturbation model in the non-regularized case $r = 1$. The horizontal axis represents the friction coefficient $c_f$ for deterministic computations and $\sigma$ for the random roughness model (blue solid line).}
  \label{fig:allFrict}
\end{figure}

Since stochastic Monte-Carlo simulations of the bottom rugosity are computationally expensive, various friction ad-hoc terms are used to model these effects. The following examples can be routinely found in the literature:
\begin{description}
  \item[Ch\'ezy law:] $\S_f = c_f\frac{u|u|}{H}$, where $c_f$ is the Ch\'ezy friction coefficient
  \item[Darcy-Weisbach law:] $\S_f = \frac{\lambda u|u|}{8H}$, where $\lambda$ is the resistance value determined according to the Colerbrook-White relation: $\frac{1}{\sqrt{\lambda}} = -2.03\log\Bigl(\frac{c_f}{14.84 H}\Bigr)$
  \item[Manning-Strickler law:] $\S_f = c_f^2\frac{u|u|}{H^{\frac43}}$, where $c_f$ is the Manning roughness coefficient.
\end{description}
The friction coefficient $c_f$ measures the bottom roughness as the parameter $\sigma$ in our random bottom roughness construction. Consequently, we can ask the same question: how does the maximum run-up value depends on the friction coefficient $c_f$ if this term is incorporated into the model? We perform a series of deterministic numerical simulations for various values of $c_f$ and the maximum wave run-up $R_{\max}$ is being measured. The numerical results are presented on Fig. \ref{fig:allFrict}. We can see that Ch\'ezy and Darcy-Weisbach laws provide a strong friction which reduces considerably the maximum run-up height. However, the Manning-Strickler law shows qualitatively a very similar behaviour to the results predicted by our stochastic model in the non-regularized case $r = 1$.

\section{Conclusions}

In the present study we considered the long wave run-up problem over rough bottoms. Specifically, we proposed a stochastic model to mimic the natural bottom roughness. Using the Monte-Carlo variance reduction technique, we quantified the maximum wave run-up behavior for various practically important values of the noise magnitude and regularity $\sigma, r$. The maximum run-up is monotonically decreasing as the bottom roughness parameter $\sigma$ increases. However, this apparent dissipative effect might be drastically reduced when the noise regularity $r$ is increased. Namely, in our simulations we observed the difference of the factor about two between the maximum run-up on the irregular ($r=1$) and regularized ($r=6$) perturbations. These results indicate that the regularity parameter has to be taken into account in some way while designing coastal protecting structures. Since the recent field survey by Fritz \emph{et al.} \cite{Fritz2007} it has been known, for example, that coastal forests do not provide effective damping to tsunamis.

Moreover, our stochastic computations were compared to several simulations using classical friction terms routinely used to model the bottom rugosity. A very good qualitative agreement (for $r=1$) was obtained with the Manning-Strickler law, while the Ch\'ezy and Darcy-Weisbach laws provide too strong momentum damping.

\section*{Acknowledgements}

D.~Dutykh acknowledges the support from French Agence Nationale de la Recherche, project MathOcean (Grant ANR-08-BLAN-0301-01) and CNRS PICS project No. 5607. The authors thank Professors W.~Craig, E.~Pelinovsky and O.~Goubet for helpful discussions.

\bibliography{biblio}
\bibliographystyle{alpha}

\end{document}